# Carbon Footprint Accounting Driven by Large Language Models and Retrieval-augmented Generation

Haijin Wang, Mianrong Zhang, Zheng Chen, Nan Shang, Shangheng Yao, Fushuan Wen, *Senior Member*, *IEEE*, Junhua Zhao, *Senior Member*, *IEEE*

*Abstract*--**Carbon footprint accounting (CFA) is crucial for quantifying greenhouse gas emissions and achieving carbon neutrality. The dynamic nature of processes, accounting rules, carbon-related policies, and energy supply structures necessitates real-time updates of CFA. Traditional life cycle assessment (LCA) methods rely heavily on human expertise, making real-time updates challenging. This paper introduces a novel approach integrating large language models (LLMs) with retrieval-augmented generation (RAG) technology to enhance the real-time, professional, and economical aspects of carbon footprint information retrieval and analysis. By leveraging LLMs' logical and language understanding abilities and RAG's efficient retrieval capabilities, the proposed method LLMs-RAG-CFA can retrieve more relevant professional information to assist LLMs, enhancing the model's generative abilities. This method offers broad professional coverage, efficient real-time carbon footprint information acquisition and accounting, and cost-effective automation without frequent LLMs' parameters updates. Experimental results across five industries—primary aluminum, lithium battery, photovoltaic, new energy vehicles, and transformers—demonstrate that the LLMs-RAG-CFA method outperforms traditional methods and other LLMs, achieving higher information retrieval rates and significantly lower information deviations and carbon footprint accounting deviations. The economically viable design utilizes RAG technology to balance real-time updates with cost-effectiveness, providing an efficient, reliable, and cost-saving solution for real-time carbon emission management, thereby enhancing environmental sustainability practices.**

## I. INTRODUCTION

ADDRESSING climate change has emerged as one of the most critical challenges of the contemporary era[1], necessitating concerted efforts to reduce greenhouse gas (GHG) emissions globally[2]. Governments worldwide have committed to ambitious dual carbon goals—carbon peaking and carbon neutrality—acknowledging the urgent need for coordinated action[3]. In this context, carbon footprint accounting (CFA) has become a pivotal tool for quantifying GHG emissions attributable to nations, industries, enterprises, and individuals[4]. Accurate CFA is essential for tracking emissions[5], assessing progress towards carbon commitments[6], and facilitating informed decision-making [7] and strategic planning[8]. Thus, the development of reliable and efficient CFA methods is of paramount importance.

Traditionally, CFA adheres to the life cycle assessment (LCA) methodology [9-13], a comprehensive approach evaluating the environmental impacts associated with all stages of a product's life cycle. LCA has become the internationally accepted method for evaluating the comprehensive environmental impact of products and technologies, extensively applied across various engineering[17] and material fields[18]. The International Organization for Standardization (ISO)[19] has established a series of standards (ISO 14040[20]/44[21]/47[22]/48[23]/49[24]/74[25]) that define the terminology, implementation, and interpretation of LCA. Building on these standards, the European Commission introduced the Product Environmental Footprint (PEF) [26] method, which encourages the calculation and disclosure of the environmental footprints of products and has piloted this approach for several typical products. Additionally, regions such as North America, Asia, and Australia have adopted similar frameworks, further promoting the application of LCA in global environmental impact assessments[27]. For instance, the United States Environmental Protection Agency (EPA)[28] and Canada's CSA Group have developed guidelines and standards aligning with ISO protocols[29], while countries like Japan and Australia have integrated LCA into their national sustainability strategies [30], emphasizing its role in reducing carbon footprints and advancing environmental sustainability.

While LCA methods offer systematic frameworks for precisely assessing environmental impacts, they depend heavily on extensive data collection and expert experience. This reliance poses challenges in achieving real-time carbon footprint accounting, particularly in reflecting real-time global production conditions. Consequently, these limitations impact strategic decisions related to carbon reduction, highlighting

---
(*Corresponding author:XXX.*)

H. Wang, M. Zhang, Z. Chen, N. Shang and S. Yao, are with with China Southern Power Grid Co., Ltd., Energy Development Research Institute, CSG, Guangdong 510663, China (e-mail: wanghj@csg.cn; zhangmr@csg.cn; chenzheng@csg.cn; shangnan@csg.cn; yaosh1@csg.cn; )

F. Wen is with Department of Electrical Engineering, Zhejiang University, Hangzhou, China. (e-mail: wenfs@hotmail.com).

J. Zhao is with School of Science and Engineering, the Chinese University of Hong Kong, Shenzhen, Shenzhen 518172, China (e-mail: zhaojunhua@cuhk.edu.cn).

the need for more efficient and real-time capable methods.

Facing the aforementioned issues, this paper introduces a novel CFA method that leverages the capabilities of large language models (LLMs) and retrieval-augmented generation (RAG) technologies to enhance the efficiency, reliability, and cost-effectiveness of real-time CFA. This approach facilitates the real-time reflection of global carbon emissions based on current production structures, aiding in timely carbon management decisions.

The contributions of this paper can be summarized as follows.

1. Enhanced Intelligent Real-Time Carbon Footprint Accounting: By combining LLMs with RAG technologies, this method significantly improves the automation and intelligence of CFA. The continuously utilize of RAG for the real-time carbon footprint datasource help obtain real-time reference carbon footprint information, optimizing prompts and enhancing the generative capabilities of LLMs. This integration accurately captures key information and subtle connections within unstructured real-time datasource, including CFA standards, accounting methods, policy documents, industry processes, and technical literature on carbon emissions across various life cycle stages. Consequently, it enable real-time information acquisition and accounting in CFA and reduce reliance on human expertise.

2. Reliable Carbon Footprint Information Retrieval: RAG technology enables the system to dynamically retrieve reference carbon footprint information from an extensive pre-constructed knowledge base, supplementing the LLMs' prompts, enhancing the reliability of LLMs. This approach ensures that the system accesses the most relevant and current professional information, which improves the overall reliability and professionalism of the CFA process.

3. Cost-Effective Carbon Footprint Accounting: Instead of frequently updating LLMs' parameters, the method combines high-frequency RAG and LLMs to establish automated carbon footprint information acquisition schemes tailored to different scenarios. This approach efficiently balances real-time updates with cost-effectiveness, ensuring that carbon footprint information can be acquired and analyzed under various conditions with minimal expense. By processing vast amounts of real-time data and continuously obtaining relevant carbon footprint fragments to optimize prompts, this method provides a low-cost solution for dynamic CFA across diverse contexts.

The rest of the paper is organized as follows. Section II provides some related work. Section III formulates the detailed method. Section IV gives the case study. Section V provides the result analysis. Finally, Section VI concludes the whole paper and gives future work.

## II. RELATED WORK AND BACKGROUND

### A. Introduction of Carbon Footprint Accounting

CFA is a crucial tool in global efforts to mitigate climate change. It quantifies GHG emissions associated with various activities, products, and processes, providing a comprehensive assessment of total GHG emissions attributable to a particular entity—be it a country, industry, organization, or individual— over a specified period. This quantification is essential for setting and achieving carbon reduction targets and formulating effective mitigation strategies.

The Life Cycle Assessment (LCA) methodology is the most widely adopted approach for carbon footprint accounting. It covers all stages of a product's life cycle, from raw material extraction, manufacturing, distribution, and use, to end-of-life treatment, known as 'cradle-to-grave'.[15] Some studies limit the scope to 'cradle-to-gate',[31] focusing on production and manufacturing processes, which are critical for raw material and energy consumption.[13]

The LCA process involves three key stages: defining system boundary, conducting life cycle inventory analysis, and performing life cycle impact assessment[14]. The system boundary definition phase establishes the study's objectives and scopes, ensuring clarity on what will be assessed and how. The life cycle inventory analysis phase involves data collection and calculation procedures to quantify relevant inputs and outputs, such as energy and raw material usage, emissions, and waste. During the life cycle impact assessment phase, the inputs and outputs are used to evaluate environmental impacts, primarily GHG emissions. These emissions are converted into carbon dioxide equivalents (CO2-eq) and aggregated to calculate the product's carbon footprint.[16]

Despite the systematic framework provided by LCA, traditional LCA-based CFA method depends on manual selection and analysis of relevant information by experts, making it difficult to scale and automate. Due to the high time and labor costs, it is challenging to achieve real-time CFA, resulting in CFA results that do not reflect real-time changes in production processes and environmental impacts.which do not reflect real-time changes in production processes and environmental impacts.

### B. Artificial Intelligence and Large Language Models

Given the challenges outlined, artificial intelligence (AI) offers promising solutions. AI, particularly LLMs, has demonstrated exceptional capabilities in understanding and generating human-like text [33], addressing some of the proposed challenges. LLMs, such as GPT-4 [34] and its successors, are AI models trained on vast datasets using advanced neural network architectures. They can perform tasks such as language translation, summarization, natural language understanding, logical reasoning, and question-answering with high accuracy and efficiency. With advanced NLP capabilities, LLMs can automate the understanding and analysis of unstructured carbon footprint datasource such as policy documents, scientific literature, and industry reports, significantly reducing the time and effort required for data collection and analysis.

Central to the success of many state-of-the-art LLMs is the Transformer architecture, introduced by Vaswani et al. in 2017.[34] Transformers have significantly improved the performance of NLP models by addressing limitations inherent in traditional recurrent neural networks (RNNs)[35] and long short-term memory networks (LSTMs)[36]. Transformers employ a self-attention mechanism that allows for greater parallelization during training, enabling the model to weigh the importance of different words in a sentence relative to one another, capturing long-range dependencies more effectively. The core structure of a Transformer consists of an encoder-decoder framework, with both the encoder and decoder built from multiple layers of self-attention and feed-forward neural networks. The encoder processes the input data to generate contextualized embeddings, while the decoder uses these embeddings to produce the output sequence. Key components of the Transformer include multi-head self-attention, which allows the model to focus on different parts of the input simultaneously, and positional encoding, which provides information about the order of words. These innovations result in state-of-the-art performance on numerous NLP benchmarks, making Transformers the backbone of many modern NLP systems. Their ability to efficiently handle large-scale data and complex language tasks underpins the effectiveness and widespread adoption of LLMs in the field.

Despite their potential in CFA, few studies have explored the application of AI for CFA. Notably, LLMs also have limitations, such as limited knowledge coverage and challenges in ensuring the factual accuracy and consistency of generated content. Additionally, the cost of updating model parameters to adapt to real-time information is high, which needs to be considered in practical applications.

This related work section now includes a comprehensive overview of current CFA methodologies and their limitations, as well as AI and LLMs advancements, specifically the Transformer architecture, addressing their potential application and limitations in the context of CFA. This provides a solid foundation for the novel CFA method proposed in this paper.

## III. PROPOSED METHOD

This study aims to design an automated CFA system based on LLMs named LLMs-RAG-CFA. This section outlines the technical roadmap and key components of the proposed method. Section III.A presents the technical roadmap of the proposed LLMs-RAG-CFA system. Sections III.B to III.D delve into the detailed mechanisms by which RAG is utilized to extract relevant carbon footprint fragments from unstructured datasource. These sections explain how the system enhances the generative capabilities of LLMs by incorporating factual knowledge retrieved through RAG, ensuring the accuracy and relevance of the generated information. Section III.E illustrates the cost-efficiency of the proposed method. Besides, the strategy employed to handle carbon footprint datasource of varying lengths is discussed. It details the different approaches adopted to optimize prompts for LLMs, balancing the need for accuracy and cost-effectiveness. This section emphasizes how the system adapts to different scenarios to maintain efficiency while reducing the overall cost of the CFA process.

### A. Roadmap

The LLMs-RAG-CFA framework mirrors the traditional expertise-based process of CFA modeling and estimation but automates key tasks that previously required extensive manual effort and expert knowledge. The framework employs LLMs and RAG to realize inventory construction and carbon emission acquisition. The system is designed to handle various scenarios, including different lengths of carbon footprint datasource, by adopting specific strategies to construct and optimize prompts fed into LLMs to obtain the necessary CFA

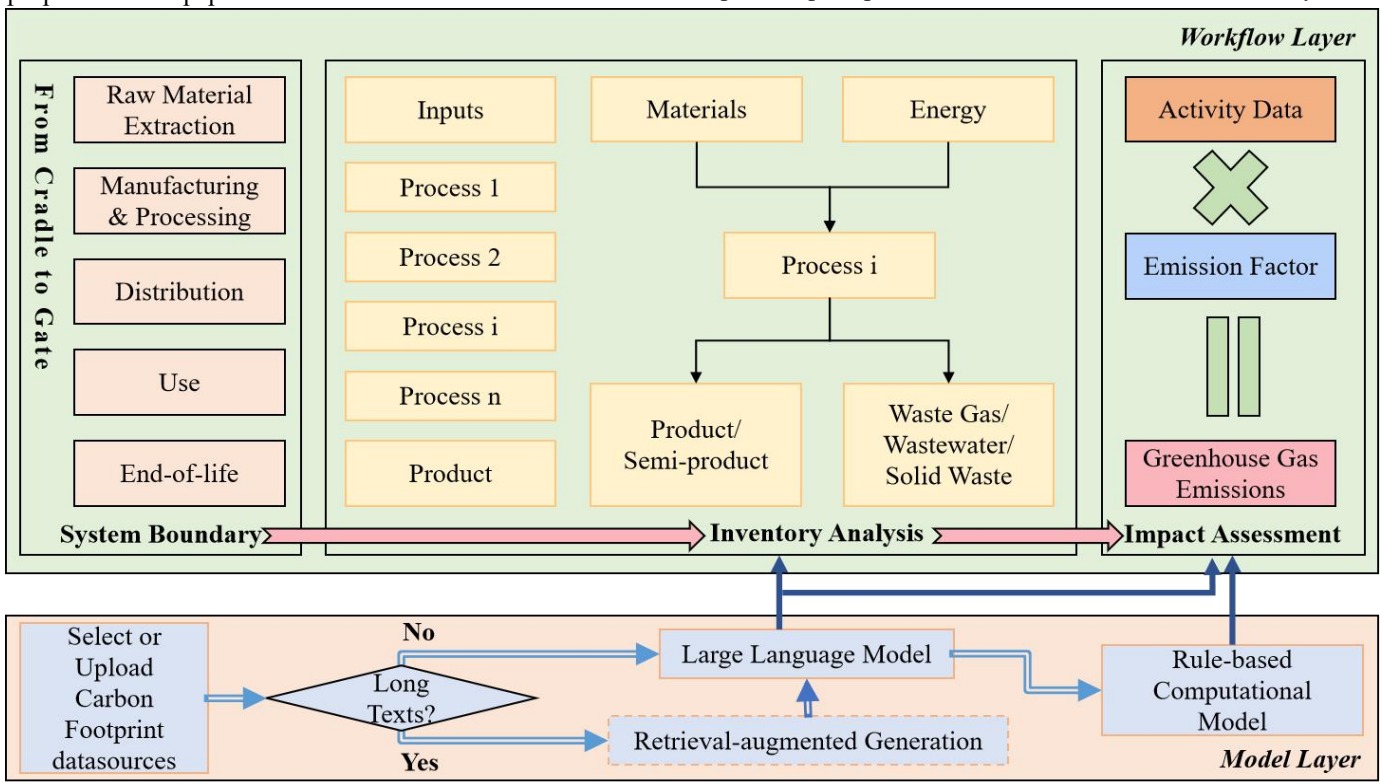

Figure 1. Architecture of LLMs-RAG-CFA

information.(Fig. 1) Then LLMs-RAG-CFA calculates the product's carbon footprint based on the inventory data and carbon emissions generated by LLMs through automated programming and encapsulation within rule-based calculation models.

LLMs-RAG-CFA leverages the power of RAG, which integrates LLMs with advanced retrieval techniques to enhance generative capabilities in CFA. The process begins by converting preprocessed and segmented carbon footprint text fragments into vectors. These vectors are then matched with user-specific carbon accounting queries through similarity calculations to retrieve the relevant carbon footprint information. This retrieval step ensures that the LLM is provided with high-quality, contextually relevant data, which it uses to generate more accurate, professional, and reliable content. The final stage involves the LLM performing knowledge fusion and answer generation, producing the comprehensive carbon footprint assessments required for CFA.

Users can initiate the CFA process by providing a prompt command that allows them to either supply a new carbon footprint datasource or select from previously stored data. Once the datasource is confirmed, users can pose specific natural language queries to obtain the necessary CFA-related carbon footprint information, including the inventory and carbon emissions with the product's production processes. The LLMs-RAG-CFA system then select the appropriate strategy based on the datasource length to processes both the user query and the provided datasource. The LLMs-RAG-CFA system processes these queries by dynamically retrieving relevant information from extensive professional data sources. Leveraging its advanced natural language understanding and generative capabilities, the system generates precise answers to the user's queries, providing the required inventory and carbon emission data. Once the relevant information is acquired, it is input into a rule-based calculation model, which then computes the final carbon footprint for the product. This approach ensures that users can efficiently and reliably obtain comprehensive and up-to-date CFA tailored to their specific needs. The framework can automatically generate the life cycle inventory and carbon footprint of the product, as well as calculate the impact assessment. This comprehensive approach significantly reduces manual workload, enhances accounting efficiency, ensures accuracy and consistency, and lowers the cost of CFA.

*B. Data Preprocessing and Text Segmentation*

The initial step involves collecting and preprocessing relevant carbon footprint data sources, which can come from user uploads or be retrieved in real-time from various carbon footprint-related websites using web scraping techniques. These sources include regulations, academic papers, industry reports, standards, and policies, which can be continuously updated to include the latest research findings and policy changes. These documents constitute a crucial data foundation to expand the professional coverage and real-time capabilities of the LLMs-RAG-CFA system.

The collected data is then categorized and converted into structured formats (e.g., JSON, XML) to ensure compatibility with machine learning models. Additionally, the data undergoes segmentation through text splitting operations, which aims to break down the documents into smaller, manageable chunks, thereby reducing subsequent computational complexity.

*C. Text Vectorization and Vector Similarity Calculation*

Text vectorization and vector similarity calculation are central to the effectiveness of the LLMs-RAG-CFA method in identifying relevant carbon footprint information.

The texts to be vectorized include carbon footprint text fragments obtained after preprocessing and segmentation, as well as user query texts. User queries are transformed into specific information needs for the carbon accounting process, such as "How much electricity is consumed per ton of primary aluminum produced smelting process?" This step aims to convert text data into high-dimensional semantic space representations, making semantically similar texts closer in vector space, while dissimilar texts are further apart. This representation facilitates accurate similarity calculation and information retrieval.

In the vectorization process, pre-trained BERT models [37] are used to convert text into high-dimensional vector representations. This involves tokenizing the input text, adding special tokens, and passing the tokenized text through embedding layers and transformer encoder layers to produce contextualized token embeddings. [38]

The training process for the vectorization model involves constructing pairs of related and unrelated carbon footprint text fragments. For example, "How much electricity is consumed per ton of primary aluminum produced smelting process?" and "What are the process parameters related to electricity consumption in the primary aluminum smelting process?" are related queries. "How much electricity is consumed per ton of primary aluminum produced smelting process?" and "What is typically chosen as the functional unit for carbon accounting of primary aluminum products?" are unrelated queries (Fig. 2). A dual-tower model (Fig.3), also known as a Siamese network [39], is employed, consisting of two identical BERT models that process the input pairs independently, producing vector representations in a shared semantic space. The objective is to minimize the distance between embeddings of related pairs and maximize the distance between embeddings of unrelated pairs. This step provides a robust data foundation for the subsequent vector similarity calculation.

The second key step is vector similarity calculation. This step measures the similarity between the user query vectors and the carbon footprint vectors to find the relevant carbon footprint text fragments. The similarity between vectors facilitates the rapid retrieval of therelevant carbon footprint text fragments. For a given user query, the top-K relevant carbon footprint text fragments are retrieved based on vector similarity measures such as dot product or cosine similarity.

Given a user query $\mathbf{q}$, the component of vector retrieval identifies the Top-K most relevant carbon footprint texts fragments $P = \{p_1, p_2, ..., p_K\}$. In detail, each query $\mathbf{q}$ and

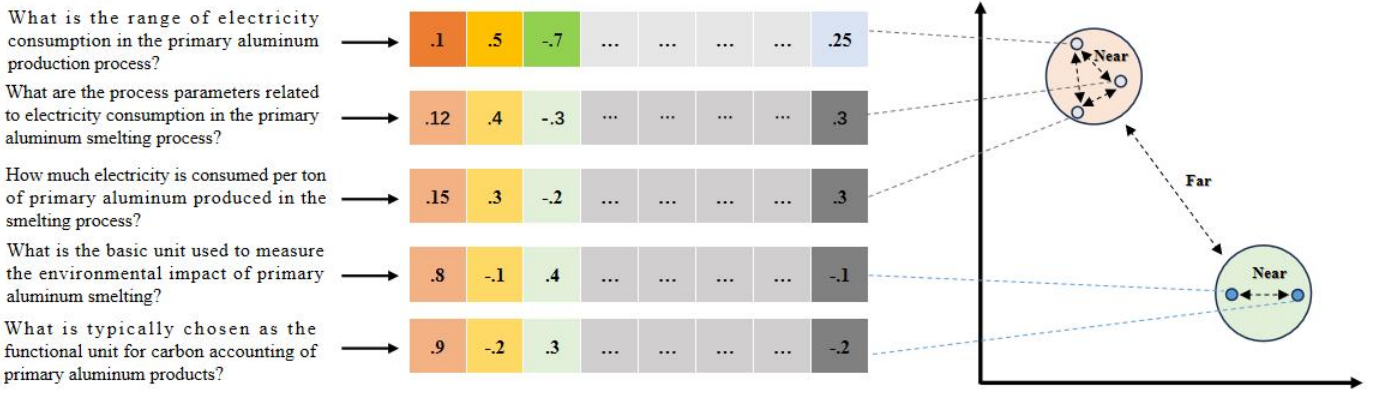

Figure 2. Vectorization example（Words with similar meanings are represented by vectors that are close in a high-dimensional space, while words with different meanings are represented by vectors that are far apart.）

carbon footprint text fragment $\boldsymbol{p_i}$ is mapped to the same semantic space vector representation:

$$\boldsymbol{v_q} = E_q(\boldsymbol{q})$$
$$\boldsymbol{v_p^i} = E_p(\boldsymbol{p_i})$$

Where $E_q$ and $E_p$ are encoder functions based on the Transformer model (BERT)[37].

The similarity $s_i$ between query vector $v_q$ and each carbon footprint fragment vector $v_p^i$ is calculated using cosine similarity:

$$s_i = sim(\boldsymbol{v_q}, \boldsymbol{v_p^i}) = \boldsymbol{v_q} \cdot \boldsymbol{v_p^i} / (\{||\boldsymbol{v_q}|| \cdot ||\boldsymbol{v_p^i}||)$$

The system then selects the Top-K carbon footprint fragments with the highest similarity as the retrieval results. This process ensures that high-quality, relevant carbon footprint information is retrieved and used to enhance the prompts fed into LLMs, leading to more accurate and professional CFA results.

### D. Knowledge Fusion and Answer Generation

In this stage, the selected top-K relevant carbon footprint text fragments, along with the user query, undergo knowledge fusion to generate an enhanced prompt, which is then processed by an LLM based on the Transformer[38] architecture. This step ensures that the prompt is enriched with pertinent details necessary for accurate carbon footprint accounting. The LLM processes the enhanced prompt, utilizing its exceptional logical understanding and generative capabilities to generate precise and contextually relevant answers to the user query.

To begin with, the input construction $\boldsymbol{x}$ involves forming the enhanced prompt by concatenating the user query $\mathbf{q}$ with the selected carbon footprint text fragments $\boldsymbol{P_{select}}$ $\{\boldsymbol{P_1}; \boldsymbol{P_2}; ...; \boldsymbol{P_K}\}$:

$$\boldsymbol{x} = [\boldsymbol{q}; \boldsymbol{P_1}; \boldsymbol{P_2}; ...; \boldsymbol{P_K}]$$

$$\boldsymbol{P_{select}} = \{\boldsymbol{P_1}, ... \boldsymbol{P_k}\} = select_k \quad (sort(\boldsymbol{S}, desc))$$
$$\boldsymbol{S} = \{s_i, i = 1, ... n\}$$

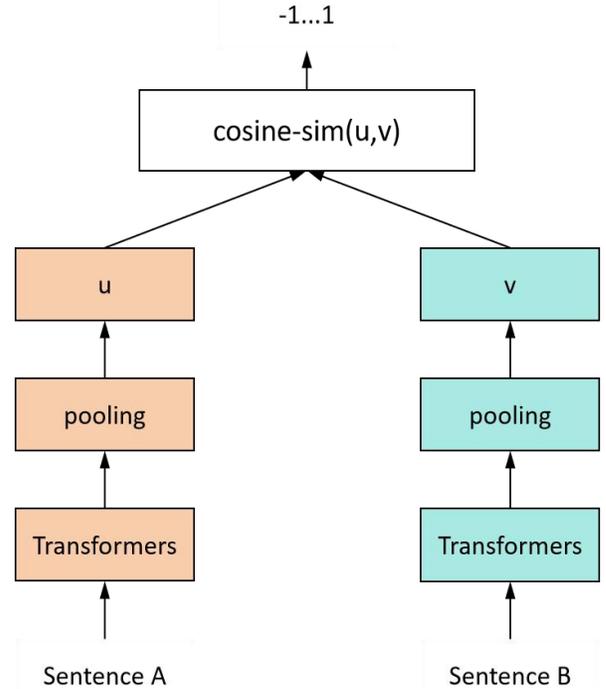

Figure 3. Dual-tower Model

Next, during the answer generation phase, the LLM leverages this enriched input $\boldsymbol{x}$ to produce the required CFA-related carbon footprint information **Output**.

In the subsequent phase of answer generation, the LLM leverages the enriched input $\boldsymbol{x}$ to produce the required CFA-related carbon footprint information, denoted as **Output.**. This process is grounded in the core architecture of the Transformer model, (Fig. 4) which is built upon a series of encoders and decoders, each composed of multiple layers. These layers include critical components such as multi-head self-attention mechanisms and feed-forward neural networks (FFN), which together enable the model to process and generate text with high accuracy.

The self-attention mechanism, which is central to the Transformer architecture, is mathematically defined as:

$$Attention(Q, K, V) = softmax\left(\frac{QK^T}{\sqrt{d_k}}\right)V$$

Where $Q$ (queries), $K$ (keys), and $V$ (values) are derived from the input embeddings. And $d_k$ is the dimensionality of the keys, which acts as a scaling factor.

The multi-head attention mechanism, an extension of the self-attention mechanism, is computed as follows:

$$MultiHead(\boldsymbol{Q}, \boldsymbol{K}, \boldsymbol{V}) = Concat(\boldsymbol{head_1}, \boldsymbol{head_2}, ..., \boldsymbol{head_h})\boldsymbol{W^O}$$

Where $\boldsymbol{head_i}$ represents the individual attention heads; $\boldsymbol{W^O}$ is a weight matrix used to transform the concatenated attention heads;

Each attention head $\boldsymbol{head_i}$ is calculated by:

$$\boldsymbol{head_i} = Attention(\boldsymbol{QW_i^Q}, \boldsymbol{KW_i^K}, \boldsymbol{VW_i^V})$$

Where $\boldsymbol{QW_i^Q}, \boldsymbol{KW_i^K}, \boldsymbol{VW_i^V}$ are learned weight matrices specific to each attention head, applied to the queries, keys, and values, respectively.

This multi-head attention mechanism allows the model to consider different aspects of the input data simultaneously, capturing a range of semantic relationships within the carbon footprint text fragments.

After the multi-head attention mechanism processes the input, the output is passed through the feed-forward neural network (FFN), defined as:

$$FFN(\boldsymbol{x}) = max(0, \boldsymbol{x}\boldsymbol{W_1} + \boldsymbol{b_1})\boldsymbol{W_2} + \boldsymbol{b_2}$$

Where $\boldsymbol{x}$ is the input vector to the FFN; $\boldsymbol{W_1}$ and $\boldsymbol{W_2}$ are weight matrices; $\boldsymbol{b_1}$ and $\boldsymbol{b_2}$ are bias terms. The function $max(0, \cdot)$ represents the ReLU (Rectified Linear Unit) activation function.

This step introduces non-linearity to the model, enhancing its representational capacity. These layers are augmented with residual connections and layer normalization:

$$\boldsymbol{Output} = LayerNorm(\boldsymbol{x} + SubLayer(\boldsymbol{x}))$$

Finally, all the generated data $\boldsymbol{Output}$ corresponding to the user query is utilized in the rule-based accounting model to obtain the result of CFA.

The effectiveness of knowledge fusion in generating enhanced prompts and feeding them into an LLM can be attributed to several key factors. Firstly, by fusing relevant carbon footprint text fragments with the user query, the enhanced prompt provides a rich context that aids the LLM in understanding the specific requirements and nuances of the query. This integration ensures that the LLM has access to all pertinent information needed to generate accurate responses. Secondly, combining multiple top-K relevant fragments into a single prompt enhances professional coverage. Each fragment contributes unique insights, covering different aspects of the carbon footprint data, leading to a more comprehensive response. Additionally, the LLM's logical understanding capabilities can uncover relationships between multiple relevant carbon footprint text fragments, resulting in tightly connected and professionally comprehensive answers. Moreover, LLMs are designed to excel at understanding and generating human-like text. Feeding them well-constructed, contextually rich prompts enables these models to leverage their deep learning capabilities fully, resulting in more precise and relevant answers. The use of top-K relevant fragments ensures that only the most pertinent information is included in the prompt, increasing the likelihood that the LLM's output will be accurate and aligned with the user's needs. This process is both automated and scalable, allowing the system to handle complex queries efficiently while reducing the manual effort required for data collection and analysis.

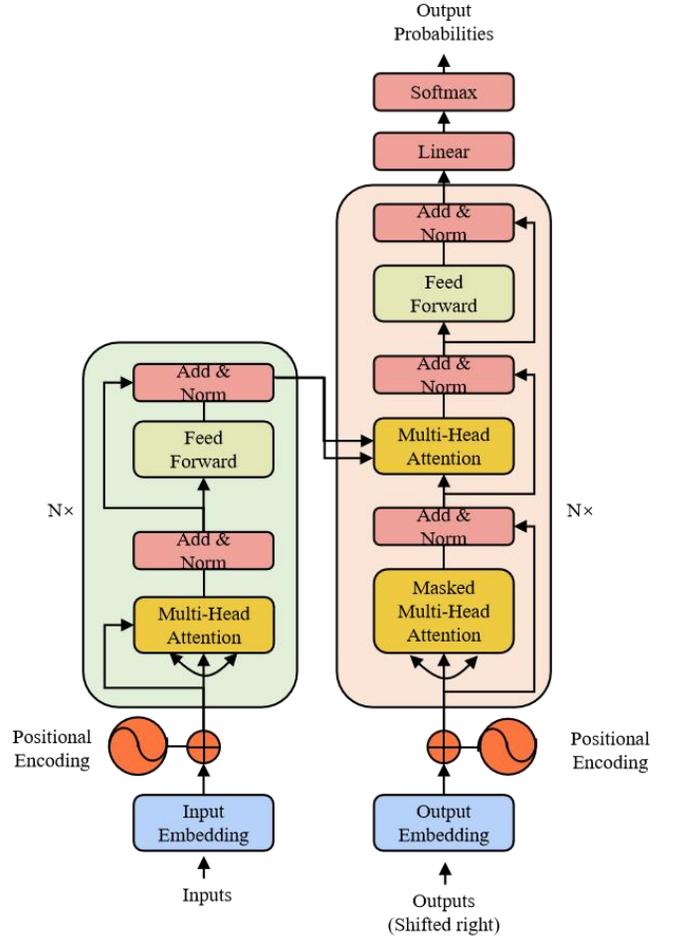

Figure 4. Architecture of Transformer

### E. Economic and Real-Time CFA based on LLM and RAG

This section aims to explore the cost-efficiency and real-time processing capabilities of LLMs-RAG-CFA.

To achieve real-time carbon footprint accounting (CFA), continuously updating LLMs' parameters to reflect the latest information can be computationally intensive and financially burdensome. To address these challenges, the RAG technology is employed in real-time to enhance the generative capabilities and professional coverage of LLMs by incorporating real-time relevant carbon footprint information. This approach significantly reduces the need for frequent LLM updates, thereby lowering computational costs while still maintaining the model's effectiveness. By generating real-time relevant carbon footprint data through RAG, the system achieves a substantial reduction in costs, coupled with improved real-time capabilities.

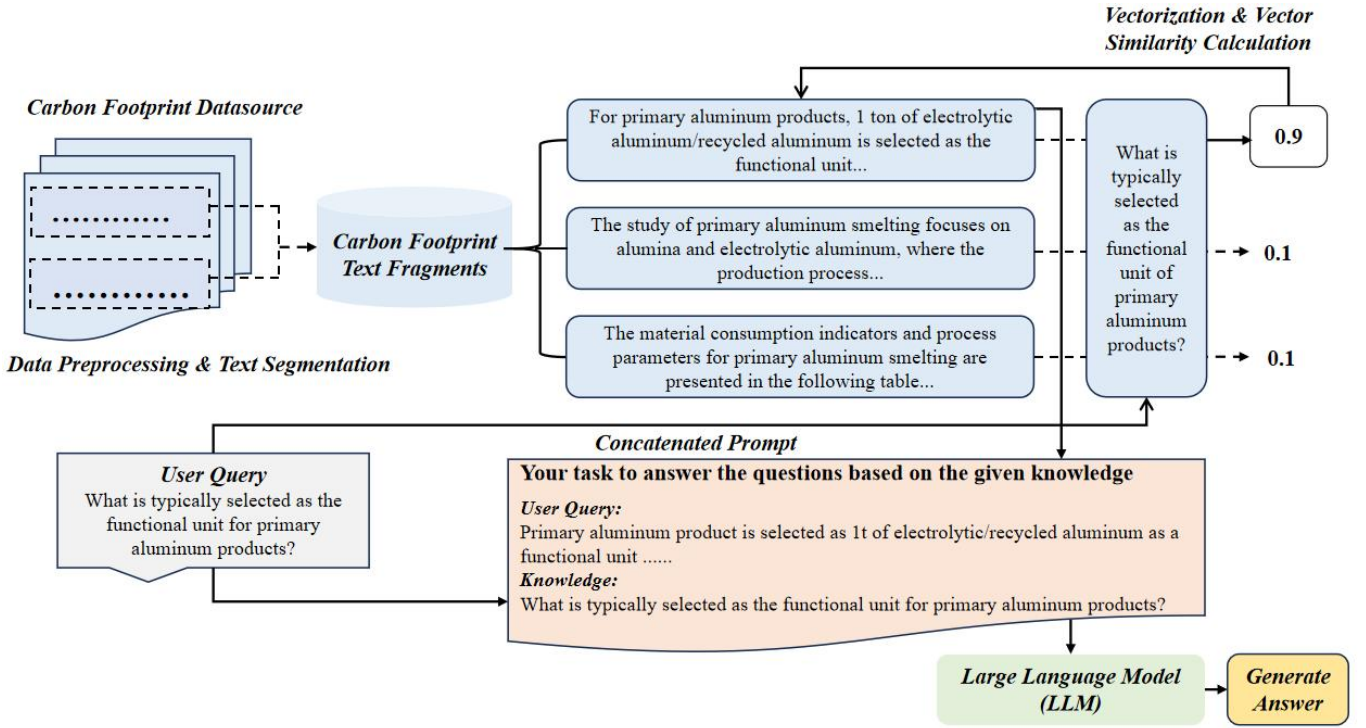

Figure 5. CFA-related carbon footprint information acquisition of LLMs-RAG-CFA when the carbon footprint datasource is provided as long texts

Additionally, different strategies are employed based on the length of the carbon footprint data source, further improving both cost-efficiency and real-time performance. As shown in Fig. 5, when handling long texts from carbon footprint data sources, the process includes several steps: preprocessing, segmentation, vectorization, and similarity calculations via the RAG methodology. This approach selects the most relevant carbon footprint text fragments, which are then used as reference information. These fragments are combined with user queries to generate enhanced prompts that are fed into LLMs, thereby augmenting their generative capabilities to produce more comprehensive and professionally accurate CFA-related carbon footprint information.

In scenarios where the carbon footprint data source is provided as short texts, these texts directly serve as reference information. The system combines the user queries with the short reference carbon footprint information to create prompts that are fed into LLMs for generating the required data. When no carbon footprint data source is provided, the system directly queries LLMs using only the user query, streamlining the processing.

Overall, LLMs-RAG-CFA improves the efficiency of CFA-related carbon footprint information retrieval and supports lifecycle inventory analysis with real-time datasource, accurately reflecting the carbon footprint of production scenarios.

## IV. CASE STUDY

### A. Scenarios and Benchmark

In this case study, CFA is evaluated across five critical industries: primary aluminum (Aluminum), lithium battery (LiB), photovoltaic (PV), new energy vehicles (NEV), and transformers. These industries significantly contribute to global carbon emissions, and their complex carbon accounting processes pose unique challenges.

To assess the performance of carbon footprint information retrieval and accounting across these industries, several methodologies were utilized as benchmarks: ERNIE Bot [40], GPT-4 [41], ChatGLM 3 [42], Qwen-14B-Chat [43] and iFlytekSpark-13B-base [44]. Within the LLMs-RAG-CFA framework, GPT-4 was employed as the LLM. These models represent a range of approaches to leveraging artificial intelligence for CFA. The carbon footprint datasource for these industries was constructed from existing literature and provided uniformly to each method. The CFA-related carbon footprint information and results generated by each method were recorded and subsequently compared against the correct answers, which were manually extracted and compiled by experts from the same carbon footprint datasource.

### B. Evaluation Index

These methodologies were evaluated based on three key metrics: carbon footprint information retrieval rate (*IRR*), information deviation (*ID*), and accounting deviation (*AD*). These metrics are essential to comprehensively assess the effectiveness of various methods in CFA.

*IRR* measures the ability of a methodology to retrieve relevant carbon footprint data from carbon footprint datasource. This metric is crucial as it reflects both the comprehensiveness of the retrieved data and the degree to which the methodology reduces reliance on manual information extraction. A higher *IRR* indicates a method's superior capability to capture a broader range of relevant data,

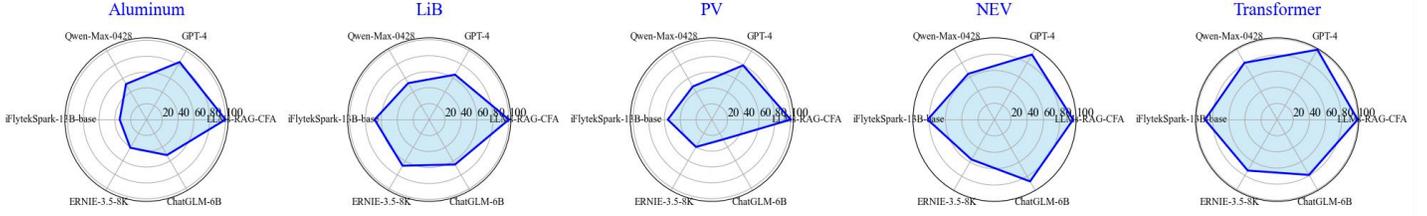

Figure 6. Available rate of information

suggesting a more effective retrieval process that leads to more thorough and professional CFA.

$$IRR = \frac{R}{T} \times 100\%$$

Where $R$ is the number of relevant carbon footprint data points retrieved, and $T$ is the total number of relevant data points available.

$ID$ quantifies the discrepancy between the retrieved carbon footprint information and the actual or true data. The true data is obtained by aggregating expert opinions based on professional literature, with outlier opinions discarded, and the average taken. This metric highlights the reliability and accuracy of the data retrieval process. The $ID$ is calculated using the Mean Absolute Percentage Error (MAPE) between the retrieved data points and the true data points:

$$ID = \frac{1}{N} \sum_{i=1}^{N} \left| \frac{D_{r,i} - D_{t,i}}{D_{t,i}} \right| \times 100\%$$

Where $N$ is the number of data points, $D_{r,i}$ is the $i$-th retrieved data point, and $D_{t,i}$ is the $i$-th true data point.

A lower $ID$ indicates that the retrieved data closely matches the true data, suggesting higher reliability and accuracy.

$AD$ measures the difference between the calculated CFA result using the retrieved data and the true carbon footprint data. This metric is vital for evaluating the overall accuracy of the carbon footprint accounting process. The $AD$ is expressed as:

$$AD = \frac{|C_r - C_t|}{C_t} \times 100\%$$

Where $C_r$ is the calculated CFA result using the retrieved data, and $C_t$ is the true carbon footprint.

A lower $AD$ signifies that the calculated carbon footprint is closer to the true value, indicating a more accurate CFA process.

These metrics provide a comprehensive framework for evaluating the effectiveness of different methodologies in CFA. By applying these metrics, the professional coverage and reliability of various approaches in CFA can be objectively assessed.

## V. RESULT ANALYSIS

### A. Information Retrieval Rate

The information retrieval rates (IRR) across various platforms for different industries are depicted in Fig. 6. The LLMs-RAG-CFA system consistently demonstrates superior performance in integrating and retrieving CFA-related carbon footprint information compared to other methodologies.

In the Primary Aluminum Industry, the LLMs-RAG-CFA system achieved a perfect IRR of 100%, significantly outperforming alternative approaches, which had IRRs ranging from 33.93% (iFlytekSpark-13B-base) to 83.93% (GPT-4). Similarly, in the Lithium Battery Industry, LLMs-RAG-CFA maintained an IRR of 100%, in stark contrast to other methods, which exhibited IRRs between 52.63% (Qwen-Max-0428) and 68.42% (iFlytekSpark-13B-base). The Photovoltaic Industry also saw LLMs-RAG-CFA demonstrate superior retrieval capabilities with an IRR of 100%, compared to other methods, which ranged from 28.97% (ChatGLM-6B) to 79.31% (GPT-4). In the New Energy Vehicle Industry, the LLMs-RAG-CFA system achieved an IRR of 96.36%, outperforming other platforms, with IRRs varying from 56.36% (ERNIE-3.5-8K) to 91.82% (GPT-4). Finally, in the Transformer Industry, LLMs-RAG-CFA maintained an IRR of 100%, surpassing other methodologies, which displayed IRRs between 72.92% (ERNIE-3.5-8K) and 100% (GPT-4).

The consistently high IRR achieved by LLMs-RAG-CFA across multiple industries indicates its superior capability in capturing a broader and more comprehensive range of relevant carbon footprint data, which is crucial for thorough and professional CFA. While pure LLMs leverage their generalist capabilities effectively, they often lack the deep, specialized knowledge required for accurate carbon footprint analysis. The LLMs-RAG-CFA system, however, bridges this gap by integrating RAG, which augments the LLMs with targeted, domain-specific information retrieval, thereby significantly enhancing the system's coverage of specialized carbon footprint data.

### B. Information Deviation Analysis

The information deviation ($ID$) across five key industries, as depicted in Fig. 7 through Fig. 11, provides insight into the accuracy of carbon footprint data retrieved by various platforms. The violin plots illustrate the distribution of ID for each method, summarizing the deviation of all CFA-related carbon footprint information across different methods within each industry. When acquiring CFA-related carbon footprint information, some LLMs and the LLMs-RAG-CFA system may output a range of values. In such cases, the ID is calculated by measuring the error for both the upper and lower boundaries of the retrieved range. The maximum absolute value of these errors is then used to represent the ID. This approach is justified as it captures the worst-case scenario in terms of accuracy, thereby providing a rigorous and conservative assessment of the method's performance.

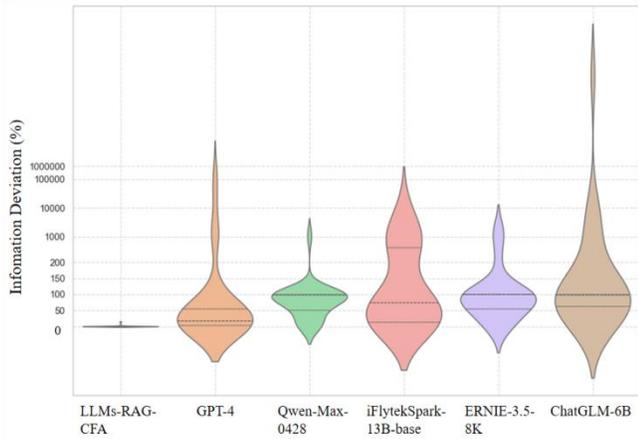

Figure 7. Information Deviation of Aluminum Industry

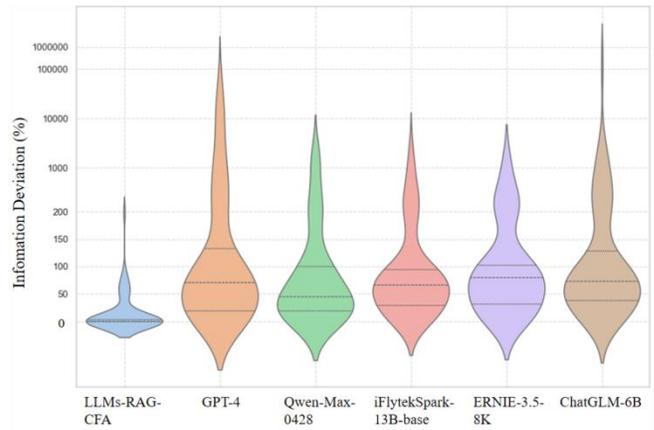

Figure 10. Information Deviation of NEV Industry

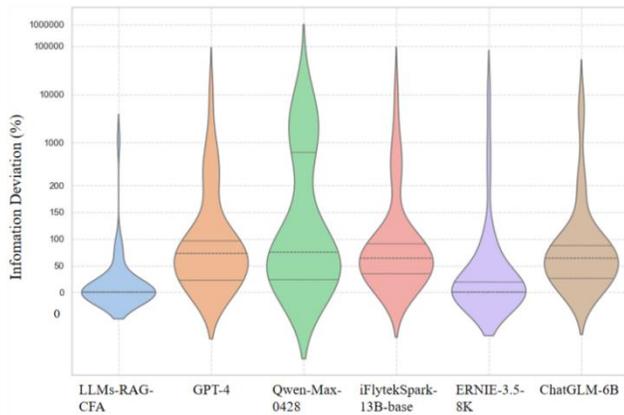

Figure 8. Information Deviation of LiB Industry

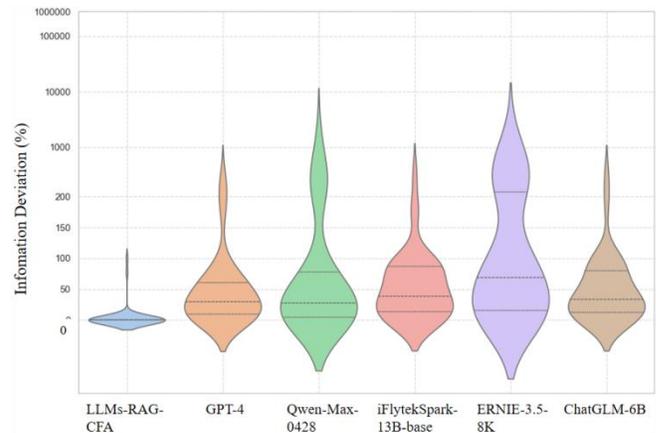

Figure 11. Information Deviation of Transformer Industry

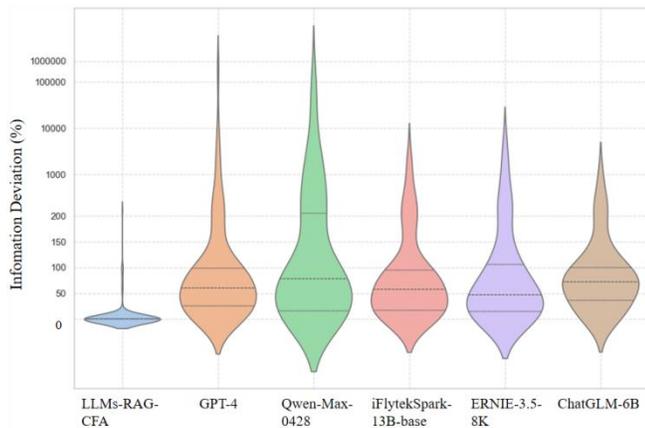

Figure 9. Information Deviation of PV Industry

In the Primary Aluminum Industry, the LLMs-RAG-CFA system exhibited a deviation range of 0% to 52.82%, markedly lower than that observed with other methods, where deviations ranged from 1.1% to 5666.67%. Similarly, in the Lithium Battery Industry, LLMs-RAG-CFA demonstrated deviations from 0% to 1465.52%, while other methods showed a broader range from 0.4% to 15017.24%. In the Photovoltaic Industry,

LLMs-RAG-CFA showed deviations from 0% to 100%, compared to the 1.1% to 68334.56% range observed in other methods. The New Energy Vehicle Industry also saw the LLMs-RAG-CFA system perform significantly better, with deviations ranging from 0% to 198.51%, whereas other methods varied between 1.91% and 171328.57%. Finally, in the Transformer Industry, LLMs-RAG-CFA exhibited a deviation range of 0% to 100%, while other methods ranged from 1.1% to 200%.

The consistently lower ID achieved by LLMs-RAG-CFA across these industries signifies its superior performance in providing accurate and reliable carbon footprint information. This reduced deviation highlights the system's ability to filter out irrelevant or incorrect data and focus on the most pertinent information for CFA. The better performance of LLMs-RAG-CFA can be attributed to the system's combination of LLMs' generalist capabilities with RAG's domain-specific information retrieval, resulting in more accurate and contextually relevant data for carbon footprint accounting. By closely approaching the information retrieval results typically achieved through manual methods, LLMs-RAG-CFA not only reduces the need for human intervention but also saves significant time and labor costs, while delivering dependable outcomes.

## C. Carbon Footprint Accounting Deviation

Fig. 12 presents the accounting deviations ($AD$) across different methods for the five industries, offering a comparative analysis of the accuracy of CFA. Similar to the ID calculation, when acquiring CFA-related carbon footprint information as a range, the AD is calculated by considering the deviations at both the upper and lower boundaries of the range. This method provides a comprehensive view of potential variations in the final CFA outcome based on the range of input data.

In the Primary Aluminum Industry, the AD for LLMs-RAG-CFA ranged between 2.35% and 19.07%, whereas other methods exhibited deviations from -206.10% to 269.36%. The reduced deviation associated with LLMs-RAG-CFA underscores its superior accuracy and reliability in carbon accounting within the aluminum industry, where precision is paramount. Similarly, in the Lithium Battery Industry, the LLMs-RAG-CFA system showed an AD of -0.07% to 0.07%, compared to -80.65% to 559% for other methods, highlighting its exceptional precision in handling complex carbon accounting scenarios. In the Photovoltaic Industry, the AD for LLMs-RAG-CFA was -1.11% to 1.11%, while other methods exhibited deviations from -92.45% to 165.80%, further illustrating its capability to maintain accuracy in rapidly evolving technological fields. The New Energy Vehicle Industry saw LLMs-RAG-CFA exhibit an AD ranging from 13.77% to -13.77%, while other methods showed deviations between 48.06% and 127.94%, reflecting the system's robustness in managing dynamic and intricate supply chains. Lastly, in the Transformer Industry, the AD for LLMs-RAG-CFA was -0.01% to 0.01%, significantly lower than the deviations observed with other methods, which ranged from -38.85% to 60.94%, emphasizing its reliability in energy management and loss calculations.

The consistently lower AD across multiple industries demonstrates that LLMs-RAG-CFA provides more accurate and reliable CFA results. This superior performance is closely tied to the system's high $IRR$ and low $ID$, which establish a more accurate data foundation for CFA. The strong performance in $IRR$ and $ID$ ensures that the LLMs-RAG-CFA system retrieves the most relevant and precise carbon footprint information, significantly reducing the errors in the final accounting results. This relationship underscores the critical role of LLMs-RAG-CFA in automating the acquisition and processing of specialized carbon footprint information, which directly enhances the accuracy and reliability of CFA outcomes. Moreover, the system's performance closely approximates that of manual methods, offering reliable results while significantly reducing human effort and time.

## VI. Conclusion

This paper introduces a novel approach to CFA that integrates LLM with RAG technology, achieving real-time CFA and significantly enhancing the efficiency, reliability, and cost-effectiveness of the process. The proposed LLMs-RAG-CFA system leverages the advanced natural language processing and logical reasoning capabilities of LLM, combined with the precise and efficient retrieval capabilities of RAG. This integration allows the system to dynamically retrieve CFA-related carbon footprint information from extensive professional carbon footprint datasource, thereby expanding the professional coverage and forming enhanced prompts to augment the generative capabilities of LLM. Additionally, the cost-efficiency and real-time processing capabilities of LLMs-RAG-CFA is illustrated. And cost-efficient mechanism is designed, ensuring that the system can generate accurate and comprehensive carbon footprint assessments with minimal expense.

Experimental results across five industries—primary aluminum, lithium battery, photovoltaic, new energy vehicles, and transformers—demonstrate that the LLMs-RAG-CFA system not only closely matches the reliability of manual methods but also significantly improves efficiency and cost-effectiveness. Compared to other LLMs, the system achieves higher information retrieval rates and significantly lower deviations in available information and CFA. And the system's economically viable design leverages RAG technology to balance real-time updates with cost-effectiveness, providing an efficient, reliable, and cost-saving solution for real-time carbon emission management.

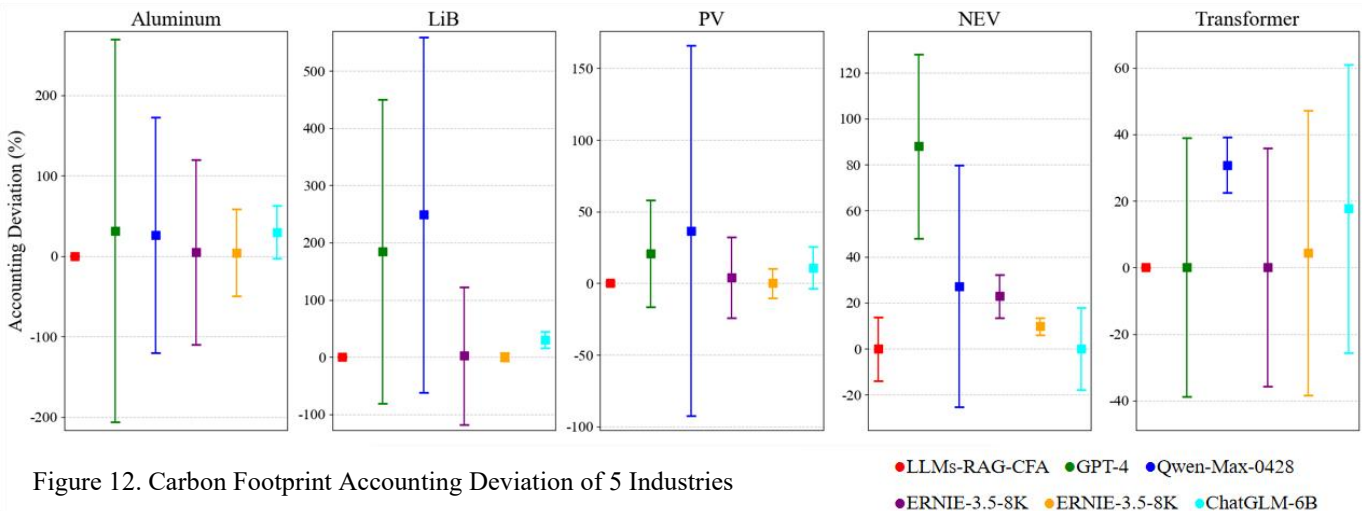

Figure 12. Carbon Footprint Accounting Deviation of 5 Industries

Looking forward, future research could explore the integration of multi-agent systems to further enhance the intelligence and adaptability of the LLMs-RAG-CFA system. Multi-agent systems can coordinate and optimize various aspects of CFA, improving real-time data processing and decision-making capabilities, thereby supporting global efforts in reducing greenhouse gas emissions.